\documentclass[twocolumn,showpacs,preprintnumbers,amsmath,amssymb]{revtex4}

\usepackage{graphicx}
\usepackage{dcolumn}
\usepackage{bm}

\begin{document}
\newcommand{\goo}{\,\raisebox{-.5ex}{$\stackrel{>}{\scriptstyle\sim}$}\,}
\newcommand{\loo}{\,\raisebox{-.5ex}{$\stackrel{<}{\scriptstyle\sim}$}\,}

\title{Nucleosynthesis of light nuclei and hypernuclei in central Au+Au collisions at 
$\sqrt{s_{NN}}$=3 GeV}

\author{ N.~Buyukcizmeci$^{1}$, T.~Reichert$^{2,3,4}$, A.S.~Botvina$^{2,3}$, 
M.~Bleicher$^{2,3,5}$}

\affiliation{$^1$Department of Physics, Sel\c{c}uk University, 42079 Kamp\"us, 
Konya, T\"urkiye}
\affiliation {$^2$Institut f\"ur Theoretische Physik, J.W. Goethe 
University, D-60438 Frankfurt am Main, Germany} 
\affiliation{$^3$ Helmholtz Research Academy Hesse for FAIR (HFHF), 
GSI Helmholtz Center, Campus Frankfurt, Max-von-Laue-Str. 12, 
60438 Frankfurt am Main, Germany}
\affiliation {$^4$ Frankfurt Institute for Advanced Studies (FIAS), 
Ruth-Moufang-Str.1, D-60438 Frankfurt am Main, Germany} 
\affiliation {$^5$GSI Helmholtz Center for Heavy Ion Research, 
Planckstr.1, Darmstadt, Germany} 

\date{\today}

\begin{abstract}
We analyze the experimental data on nuclei and hypernuclei 
yields recently obtained by the STAR collaboration. The hybrid 
dynamical and statistical approaches which have been developed 
previously are able to describe the experimental data 
reasonably. We discuss the intriguing difference between the 
yields of normal nuclei and hypernuclei which may be related to 
the properties of hypermatter at subnuclear densities. 
New (hyper-)nuclei could be detected via particle 
correlations. Such measurements are important to pin down 
the production mechanism. 
\end{abstract}


\maketitle


\section{Introduction} 

During recent years the production of new nuclei has become again 
one of the central topics in relativistic nuclear reaction studies. 
It is known since the late 1970s that many different light 
complex nuclei can be formed in central nucleus-nucleus collisions 
\cite{Gos77}. Later on these studies were considerably extended and 
presently they involve the production of both normal nuclei and 
hypernuclei, including exotic nuclear species. In central relativistic 
nucleus-nucleus collisions the yields and spectra of hydrogen and helium 
isotopes have been observed. In addition, more heavy species, like 
Li, Be and others were also under examination 
\cite{FOPI97,Neu00,Neu03,FOPI10}.
It is commonly accepted these light nuclei are mostly formed on later stages 
of the reaction from nucleons which are primary produced. Although other 
production mechanisms, like direct thermal production, have been proposed 
\cite{And11}.  
This indicates that nucleosynthesis mechanisms can be studied in 
these processes and they may complement and lead to better understanding 
the nucleosynthesis in the Universe. Recently the detection of 
hypernuclei was also reported in relativistic nuclear collisions 
\cite{alice13,ALICE,STAR,HADES} 
providing an opportunity to extend the nucleosynthesis investigation to 
strange nuclear matter. Also heavy-ion collisions open new possibilities to 
obtain novel hypernuclei, including multi-strange and exotic hypernuclei, 
which are more difficult to produce in other reactions. 

In this paper we further develop the theory of hypernuclei production 
by analyzing the formation of light nuclei and hypernuclei which were 
measured at the STAR experiments with fixed targets in gold+gold 
collisions \cite{STAR}. As we have previously demonstrated in our calculations 
the reactions with a beam energy around 3--10 GeV per nucleon lead to 
a large production of hyperons and can provide 
essential yields of hypernuclei \cite{Bot13,Bot17}. Therefore, the systematic 
comparison with experiments in this energy range is important for both 
central and peripheral collisions. Recently we have succeeded to describe 
the formation of complex normal nuclei in central collisions \cite{Bot21,Bot22}. 
Now we demonstrate that the same nucleosynthesis mechanism can be 
effectively applied to describe the hypernuclear production too.

\section{Mechanisms of the nuclei production}

In the most general consideration the reaction may be 
subdivided into several stages: (1) The dynamical stage which can lead to 
the formation of an equilibrated nuclear system. (2) The statistical 
fragmentation of such system into individual fragments, 
which can be accompanied by the de-excitation of these hot fragments if 
they are in an excited state. Various transport models are currently 
used for the description of the dynamical stage of the nuclear reaction 
at high energies. They take into account the hadron-hadron interactions 
including the secondary interactions and the decay of hadron resonances 
(e.g. \cite{Aic91,Ble99}). For this reason they preserve important 
correlations between hadrons originating from the primary interactions in 
each event, which are ignored when we consider the final inclusive 
particle spectra only. Using dynamical models it was established that many 
particles are involved in these processes by the intensive rescattering leading 
to collective behaviour during the evolution at the collisions. In peripheral 
collisions the produced high energy particles leave the system and the 
remaining nucleons form an excited system (a residue). We may expect 
that this system evolves toward a state which is mostly determined by 
the statistical properties of the excited nuclear matter. Its decay leads 
to the production of various new nuclei (see, e.g., \cite{SMM,Ogu11,EOS}). 

It is typical for relativistic collisions of two nuclei that  as a result of the 
dynamical production of individual baryons a substantial amount of hyperons 
and nucleons populate the midrapidity kinematic region.  In this region new 
nuclei can be formed  from these particles via their subsequent interaction 
in the diluted nuclear matter during the expansion process. Generally, 
at the end of the dynamical stage (at a time around $\sim$20-40 fm/c after 
the beginning of the collision) many new-born baryons escape from 
the colliding nuclei remnants. Some of these baryons may be located in the 
vicinity of each other with local subnuclear densities around 
$\sim$0.1$\rho_0$ ($\rho_0 \approx 0.15$ fm$^{-3}$ being the ground 
state nuclear density). This nuclear matter density corresponds to the 
coexistence region in the nuclear liquid-gas type phase transition. 
It is the proper place of the synthesis of new nuclei, because the remaining 
attraction between baryons can lead to complex fragment formation. 
We expect that such nucleation processes will mostly produce the light 
nuclei. The formation process may be simulated as the baryon attraction 
using potentials within the dynamical transport 
models \cite{FRI,Aic20,Gla22}, or within phenomenological coalescence models 
\cite{Ton83,Neu00,Bot15,Bot17a,Som19}. However, as we have demonstrated 
recently \cite{Bot21,Bot22}, the clusterization processes can be reasonably 
described as the statistical formation of nuclei in the low density matter 
in local chemical equilibrium. 

To describe the dynamical reaction part we use  the transport 
UrQMD model \cite{Ble99,urqmd1}, which is modified for the 
present studies. UrQMD is  quite successful in the description of 
a large body of experimental data on particle production \cite{Tom22,Tom23}. 
The current version of UrQMD include up 
to 70 baryonic species (including their antiparticles), as well as up 
to 40 different mesonic species, which participate in binary interactions. 
In the present calculations the hard Skyrme type equation of state is used 
which includes also  for the attraction between baryons. We explicitly 
conserve the net-baryon number, net-electric-charge, and net-strangeness as 
well as the total energy and momentum.  The produced particles can be 
located at various rapidities, however, the main part is concentrated 
in the midrapidity region. After a time of 20--40 fm/c the strong interactions 
leading to the new particle formation cease and the system start to decouple. 
Such kind of a freeze-out is a general feature captures in the 
transport approaches \cite{Bot11}. 
In that time-moment we consider the relative coordinate positions and 
velocities of the produced baryons. We select the nuclear clusters according 
to the coordinates and velocities proximity, as was suggested in 
Refs.~\cite{Bot15,Bot21}, and we call it clusterization of baryons (CB). 

In our CB procedure the diluted nuclear matter 
is subdivided into many clusters with a coalescence-like recipe. 
In particular, we assume that baryons (both nucleons and hyperons) can 
produce a cluster with mass number $A$ if their velocities relative to the 
center-of-mass velocity of the cluster is less than a critical velocity $v_c$. 
Accordingly we 
require $|\vec{v}_{i}-\vec{v}_{cm}|<v_{c}$ for all $i=1,...,A$, where 
$\vec{v}_{cm}=\frac{1}{E_A}\sum_{i=1}^{A}\vec{p}_{i}$ ($\vec{p}_{i}$ are 
momenta and $E_A$ is the summed energy of the baryons in the cluster). 
In addition, we assume that the distance between the individual baryons 
and the center of mass of the clusters should be less than $2\cdot A^{1/3}$ fm, 
so these baryons can still interact leading to the nuclei formation.
In this case such clusters with nucleons inside have the density of 
$\rho_c \approx \frac{1}{6} \rho_0$ as it was established in the previous 
studies of the statistical multifragmentation process 
\cite{SMM,MMMC,Xi97,MSU,INDRA,TAMU,Vio01,FASA}. 
Since the baryons do still move with 
respect to each other inside these clusters, 
they present an excited nuclear system. 
The excitation energy of such clusters is calculated according to the 
method suggested in Refs.~\cite{Bot21,Bot22}. 
The excitation energy of the 
clusters is related to the properties of nuclear matter in local equilibrium 
at the cluster density. It is also connected to the corresponding baryon 
interaction in matter. 
These clusters are analogous to the local freeze-out states for 
the liquid-gas type phase coexistence adopted in statistical models. 
The following evolution of the clusters, 
including the formation of nuclei from these baryons, can be described in 
a statistical way. 
According to our procedure these hot clusters decay into nuclei. 
For the description of this process we employ the Statistical 
Multifragmentation 
Model (SMM) which describes the production of normal nuclei very well, and 
it was generalized for the hypernuclear case 
(see Refs.~\cite{SMM,Ogu11,EOS,Bot07,lorente}). 

Within this approach we have succeeded to describe the yields and energies 
of light nuclei observed in the FOPI experiment for central collisions of 
relativistic heavy ions \cite{FOPI97,FOPI10}, 
that was not possible with the previous models. 
It is especially important since we have explained the cross-over behaviour 
of the $^3$He and $^4$He light nuclei production with the beam energy: 
The $^4$He yield dominates over $^3$He at low and intermediate energies, 
while the $^3$He yield is larger at high energies. This was previously not 
understandable within simplistic coalescence models which always produce 
more light nuclei than heavy ones. The reason of this behaviour is in the 
decay of the primary large clusters that favor the $^4$He production in 
comparison with $^3$He. For high collision energies the sizes of the primary 
clusters becomes smaller and leads after their decay 
to the production of smaller nuclei. The important result of our studies is 
that the excitation energy of such clusters should not be too high, 
around $\sim 10$ MeV per nucleon, i.e., close to the nuclear binding 
energy, and corresponding to the coexistence region of the nuclear
liquid-gas phase transition. Based on these previous successes of our approach we extend it now 
to analyze the hypernuclear observations.

\section{Excited nuclear clusters at subnuclear density}

\begin{figure}[tbh] 
\includegraphics[width=8.5cm,height=13cm]{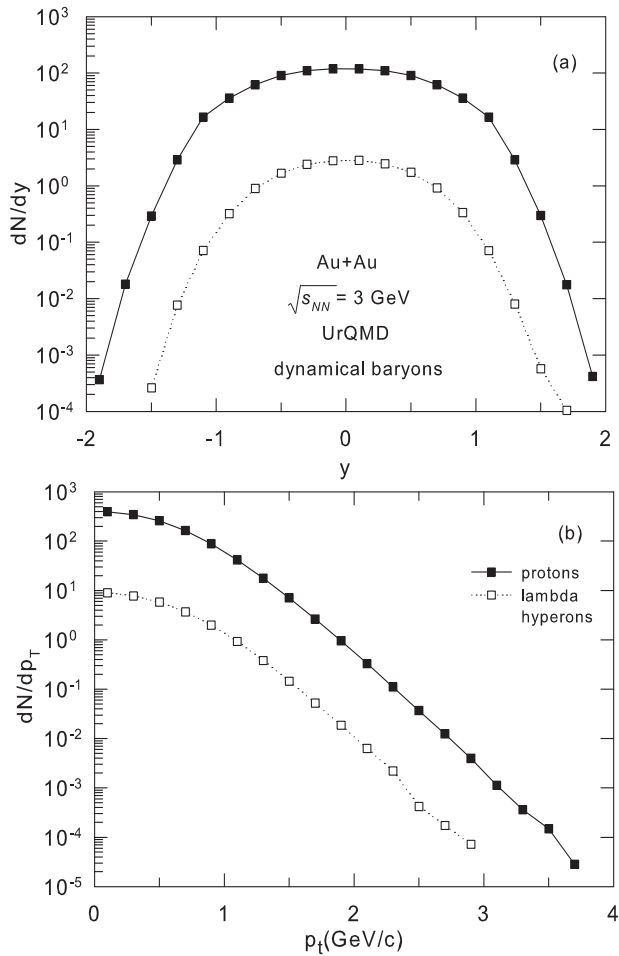} 
\caption{\small{ 
Total proton and $\Lambda$ distributions (per event) after UrQMD calculations 
of central gold collisions at center-of-mass energy of $\sqrt{s_{NN}}$=3 GeV. 
Top panel (a) - rapidity distributions. 
Bottom panel (b) -  transverse momenta distributions, in the rapidity range 
$|y| < 0.5$. 
}} 
\label{fig1} 
\end{figure} 


In Fig.~\ref{fig1} we show the distributions of protons and Lambda hyperons after 
the UrQMD simulations of central Au~+~Au collisions (impact parameters 
$b \leq 3$~fm) 
at $\sqrt{s_{NN}}$=3 GeV center-of-mass energy. (The considered $\Lambda$ 
do not include $\Sigma^0$ hyperons since they decay later than the hypernuclei 
are formed.) One can see that the baryons have 
very broad momentum distributions going beyond the projectile and target rapidities. 
We have also performed the calculations with other models, the Dubna cascade model 
\cite{Ton83,Bot21} and the phase space generation of all available particles in 
the center of mass system \cite{Bot22}. The results are very similar, therefore, 
we are sure to reproduce the general reaction picture. 
These distributions characterize the baryon UrQMD input 
for the following cluster selection. To evaluate the dynamical uncertainties the 
two time instances, 20 fm/c and 40 fm/c after the collision, are analyzed. Since 
the interaction rate decreases rapidly the momentum distributions change very 
little during the times. The coordinate distances between baryons increase towards 
the later time. 
Nevertheless, it has a little effect on the following clusterization in the CB procedure 
in the case when we consider $v_c$ parameters (0.14$c$ and 0.22$c$) which were 
previously extracted as the best ones from comparison with FOPI experimental 
data \cite{FOPI97,FOPI10}. Such a low sensitivity is because 
the hadron correlations are propagated explicitly in UrQMD. 

\begin{figure}[tbh] 
\includegraphics[width=8.5cm,height=13cm]{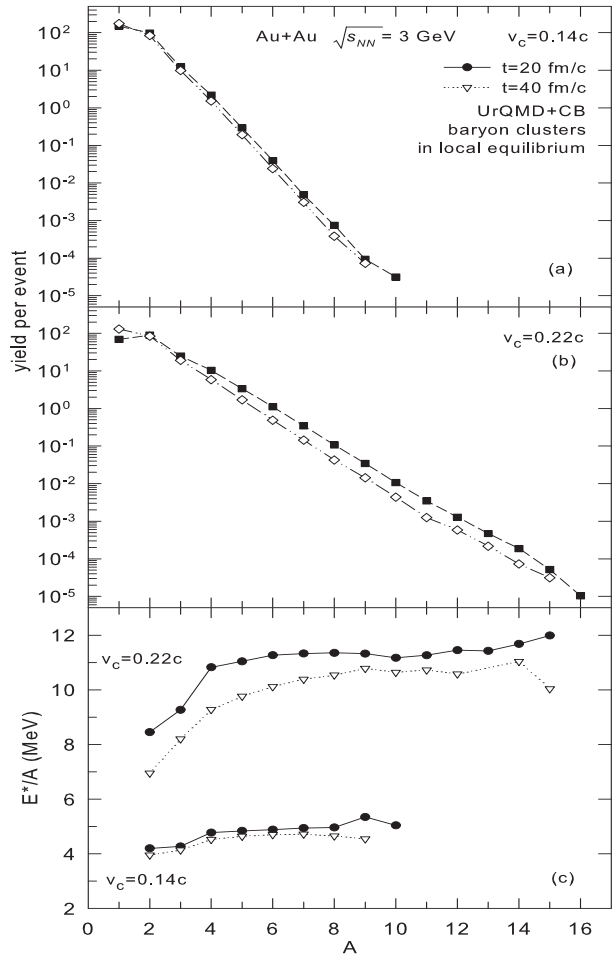} 
\caption{\small{ 
Calculated distributions of local nuclear clusters (per event) 
formed from dynamically produced baryons after UrQMD and the 
clusterization (CB) procedure by using the selection of the  
baryons with the velocity and coordinate proximity. 
Top panel (a) - mass distributions of the cluster with the 
velocity parameter $v_c$=0.14$c$. 
Middle panel (b) - mass distributions of the cluster with the 
velocity parameter $v_c$=0.22$c$. 
Bottom panel (c) - average excitation energy of the clusters versus 
their mass number. The times for stopping the UrQMD calculations and 
$v_c$ parameters are shown in the panels. 
}} 
\label{fig2} 
\end{figure} 


The results of the following selection of the baryonic clusters within CB are shown 
in Fig.~\ref{fig2}. We  see that the cluster yields decrease nearly 
exponentially with their masses, and it is similar to a normal coalescence-like 
process. As was mentioned, in our approach we assume that these clusters 
are excited nuclear systems in local chemical equilibrium. The bigger clusters 
can naturally be formed with larger velocities parameters  $v_c$. 
The time dependence of the cluster sizes is also understandable: 
It is obvious that the larger coordinate distances between baryons slightly decrease 
the production of big clusters because of the larger space separation. 
The important characteristic is the excitation energy of these clusters, which are 
presented in the bottom panels. A larger $v_c$ leads to a higher 
excitation energy, since the relative velocities of the baryons inside the clusters are 
higher. 

\begin{figure}[tbh] 
\includegraphics[width=8.5cm,height=13cm]{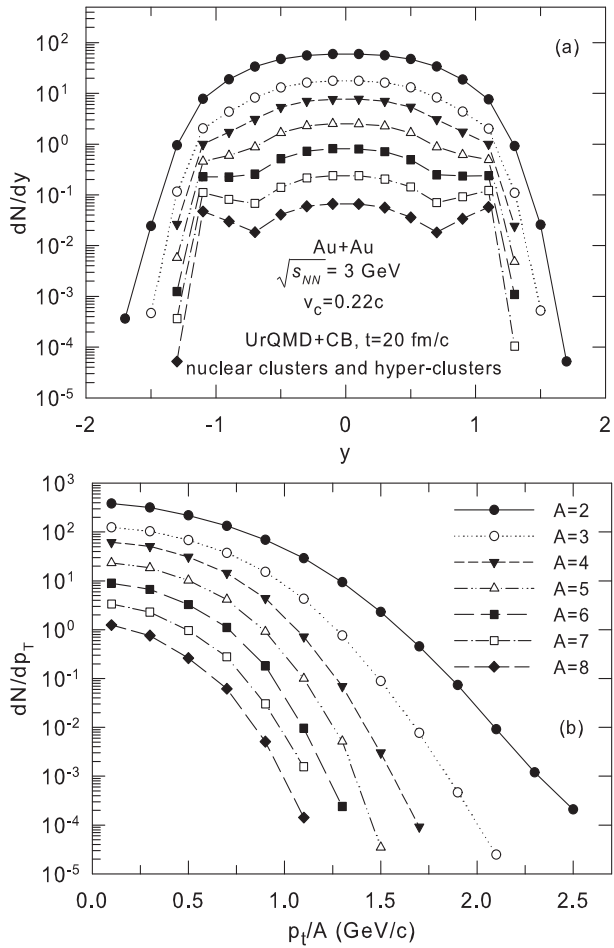} 
\caption{\small{ 
Rapidity (top panel (a)) and transverse momentum per nucleon (bottom panel (b) 
overall rapidities) distributions of the excited baryon clusters, which have 
the mass numbers from $A$=2 to 8. Yields are per event. 
}} 
\label{fig3} 
\end{figure} 

 
The kinematic characteristics of the primary excited clusters (rapidity and 
transverse momenta) are depicted in  Fig.~\ref{fig3}, for the case 
of $v_c$=0.22$c$. Reflecting the properties of the initial baryons (Fig.~\ref{fig1}) 
these rapidity and transverse distributions are quite broad. It is interesting 
that relatively large clusters can be eventually formed
both in the midrapidity region and the target/projectile kinematic region.
This is a consequence of that even in central collisions some nucleons of
the target and projectile may go through the nuclear matter with a small
probability. Further information 
about new nuclei after their production via the decay of such clusters one can 
find in Refs.~\cite{Bot21,Bot22}. For example, the products of the cluster decay 
will preserve the kinematic characteristics (per nucleon) corresponding 
to the dynamically produced baryons. Many new 
exotic nuclei can be formed, and the specific particle correlations are the 
best way to distinguish this reaction mechanism.

\section{Analysis and interpretation of experimental data}

\begin{figure}[tbh] 
\includegraphics[width=8.5cm,height=13cm]{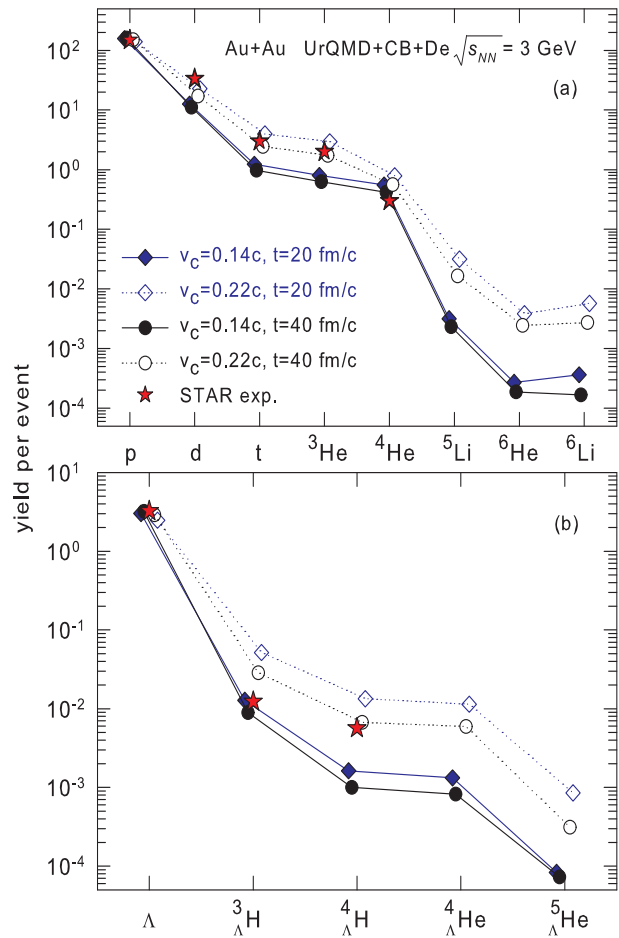} 
\caption{\small{ 
(Color online.) 
Comparison of full calculations including UrQMD transport, formation 
of excited local thermalized clusters (CB) and de-excitation of these 
clusters (De) with STAR experimental data. The predictions for other 
important nuclei and hypernuclei are presented too.
Top panel (a) - total yields of normal nuclei in central collisions. 
Bottom panel (b) -  total yields of hypernuclei in central collisions. 
Notations for nuclei, and the used time and $v_c$ parameters are shown in 
the panels. 
}} 
\label{fig4} 
\end{figure} 

\begin{figure}[tbh] 
\includegraphics[width=8.5cm,height=13cm]{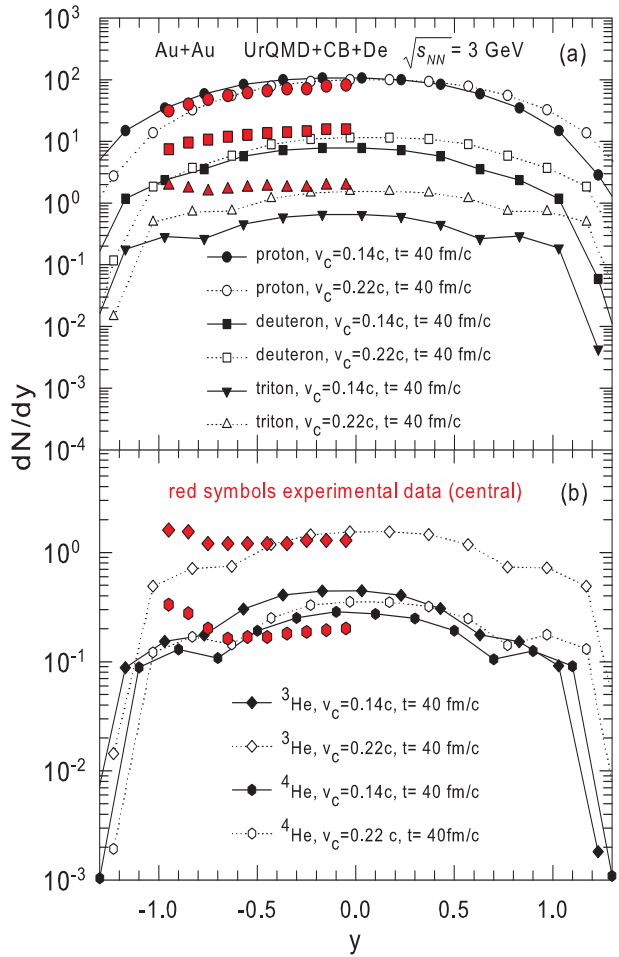} 
\caption{\small{ 
(Color online.) 
Comparison of the calculations of the rapidity distributions of normal 
nuclei produced in central collisions with STAR experimental data 
\cite{STAR1}. 
Top panel (a) - protons, deutrons, tritons. 
Bottom panel (b) - $^3$He, $^4$He.  
The model parameters are indicated in the panels.
}} 
\label{fig5} 
\end{figure}
\begin{figure}[tbh] 
\includegraphics[width=8.5cm,height=13cm]{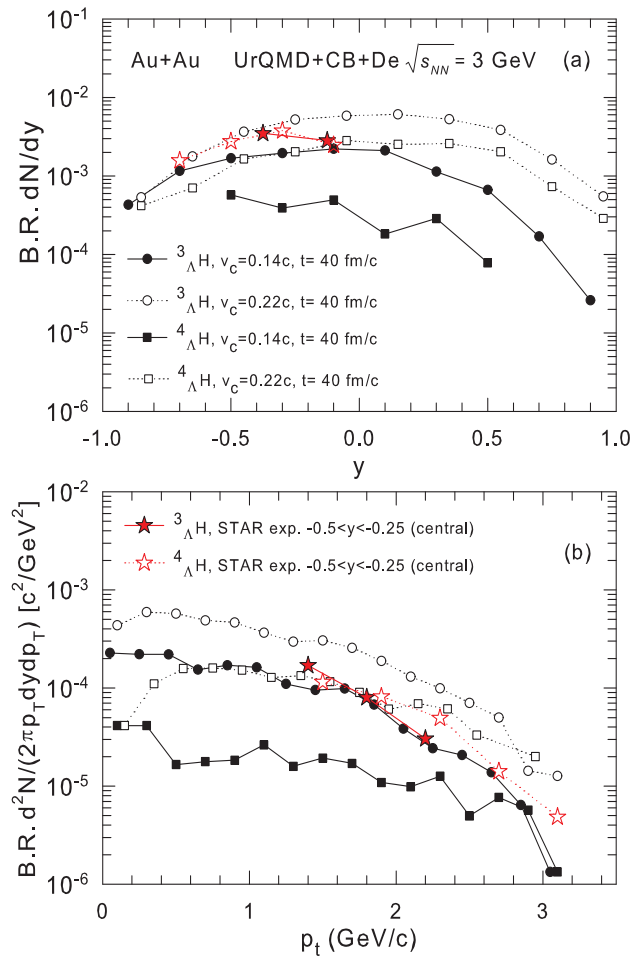} 
\caption{\small{ 
(Color online.) 
Comparison of the calculations of the rapidity (top panel (a)) and transverse 
momentum distributions (bottom panel (b)) of $^3_{\Lambda}$H) and 
$^4_{\Lambda}$H) hypernuclei with STAR experimental data \cite{STAR}. 
The model parameters and the rapidity intervals are shown in the panels. 
The branching decay ratios (B.R.) are taken into account in the calculations. 
}} 
\label{fig6} 
\end{figure}

In this paper we concentrate only on the analysis of the recent STAR data. 
In central collisions the main contribution to the production of
hypernuclei comes after the secondary interaction of baryons inside
the excited nuclear clusters which have subnuclear densities and
temperatures corresponding to the phase coexistence. As a part of
microcanonical SMM describing this process as the cluster decay we
have involved the generalized statistical Fermi-break-up
model \cite{lorente}. In this way we can investigate the properties
of the excited hyper-matter via its disintegration. For example,
by comparing the yields of different hypernuclei one can get
information about their binding energy and hyperon interaction in
the matter \cite{Buy18}. The comparison with the experimental data on nuclei and hypernuclei 
production \cite{STAR,STAR1} and the predictions within our approach are 
shown in Figures~\ref{fig4},~\ref{fig5}, and~\ref{fig6}. 
We are able to reproduce the main experimental results (involving the 
distributions trends) by applying the models discussed in the previous 
sections with reasonable parameters. Therefore, we believe that the idea 
including the formation of the baryonic clusters under local equilibrium 
and their following statistical disintegration is very promising. 

In Fig.~\ref{fig4} we show the yields of some light nuclei and hypernuclei 
(normalized per event) obtained in central collisions. In the STAR experiment 
these yields were evaluated with a special procedure, however, we do it on 
an event per event basis directly from the calculations. We can reach the best 
description of the data for normal fragments with the mass numbers $A=2-4$ 
by using the parameter $v_c$=0.22$c$ . The yields manifest the exponential 
decreasing with $A$, that is expected from other experiments too. In our 
calculations the final nuclei with $A \sim 4 $ we have obtained after the 
de-excitation of local primary clusters with $A \sim 10$ 
and with the corresponding excitation energy of around 10 MeV 
per nucleon (see Fig.~\ref{fig2}). This excitation energy is consistent with 
the one extracted from the analyses of other experiments \cite{Bot21,Bot22}. 

The rapidity distributions of produced normal light nuclei are depicted in 
Fig.~\ref{fig5}. Our calculations are in qualitative agreement with the data. 
However, the experimental yields of nuclei for larger species show a 
slight increase towards the target rapidity. We believe it is related to the 
experimental selection of central events (0--10\%) via the particle multiplicity 
and the applied spectator cut. In reality large multiplicities may be also 
obtained in quasi-peripheral 
events, when a large group of nucleons is still located in the target and 
projectile kinematic regions. These nuclear remnants are excited, and they may 
even capture hyperons \cite{Bot15,Bot17}. Namely the decay of such residues 
can give an additional contribution to the yields of larger nuclei. The goal of 
the present calculations is to investigate the true central events ($b \leq 3$~fm) 
without this contribution.

Fig.~\ref{fig6} shows the rapidity and transverse momentum distributions 
of the hypernuclei  ($^4_{\Lambda}$H and $^3_{\Lambda}$H). We are able 
to reproduce correctly the main trends of the distributions. With $v_c$ 
parameters which are in the range suitable for the description of normal nuclei 
we may describe the total yield of one of these two hypernuclei. However, we see 
that in the experiment the production of $^4_{\Lambda}$H nuclei looks similar 
to $^3_{\Lambda}$H nuclei before the branching ratio correction. While in the 
calculations it is essentially lower, even by taking the corrections into account. 

It is also important that by examining carefully Fig.~\ref{fig4} we see an 
essential difference between the experimental yields of hypernuclei and normal 
nuclei, which are obtained after all experimental corrections. Namely, the yield 
ratio of $^3_{\Lambda}$H to $^4_{\Lambda}$H is less than a factor 2. Whereas 
the ratio of $^3$He to $^4$He is more than factor 5. To our opinion it is one of 
the most interesting and puzzling results of this experiment: 
The general decrease of the yields with mass number $A$ is a quite expected 
behaviour and it is confirmed by many other experimental measurements. 
Phenomenologically one can understand it as a low probability to add an 
additional nucleon to the formed fragments. However, the data seems to 
suggest that an increase by one neutron for $A=3$ fragments has a much 
larger probability in the hypernuclear case as in the case of a normal nucleus. 
We note that our results were obtained by including in the statistical
model the excited states of hypernuclei which importance for its
production was previously widely discussed, see e.g.
Refs.~\cite{lorente,Leun23}. 

We think this effect may be used extract an important experimental 
information on the properties of hypermatter of subnuclear density and on 
the $\Lambda$ interaction in this matter. 
We have investigated possible reasons of this effect within our approach. 
It is known that the $\Lambda$ interaction in nuclei is lower than for 
nucleons approximately by factor 2/3. One can phenomenologically suppose 
that it is related to the number of $u$ and $d$ quarks in the baryons. 
For this reason the excitation energy of the primary local hyper clusters 
may be lower than in normal clusters. We have performed the corresponding 
calculations by assuming that the excitation of such clusters are only 2/3 of 
the normal clusters to simulate the effects of the hyperon 
interaction in the low-density clusters. In the same time the properties 
of normal clusters dominating in the system remain unchanged. 
We obtain that the yields of $^4_{\Lambda}$H increases much 
stronger than $^3_{\Lambda}$H, and the ratio becomes closer to the 
experimental one. 
Another possibility to explain this effect is related to the prolonged space 
structure of weakly bound $^3_{\Lambda}$H nucleus 
formed in the phase transition coexistence region. In the statistical model 
it can be taken into account by reducing the coordinate phase 
space of such hypernuclei, as it was suggested before for some normal 
nuclei \cite{SMM}. This can decrease the calculated $^3_{\Lambda}$H 
yield. 

Presently we do not have a sufficient experimental information to 
distinguish between the two alternative scenarios. 
In particular, the production of other nuclei and hypernuclei in same 
reactions is crucially important for the understanding of the underlying 
physical process. In Fig.~\ref{fig4} we show the predictions for the 
yields of few heavier nuclei ($^5$Li, $^6$He, $^6$Li), and hypernuclei 
($^4_{\Lambda}$He, $^5_{\Lambda}$He), which should be produced in 
addition within the corresponding reaction mechanism. 
The $^5$Li production can be measured via its decay into proton and 
$^4$He, or into $^2$H and $^3$He, with a probability compared with 
the probability of correlations measured for identification of 
$^3_{\Lambda}$H and $^4_{\Lambda}$H. The $^4_{\Lambda}$He and 
$^5_{\Lambda}$He hypernuclei can be identified via three particle 
correlations. Generally, more heavy nuclei and hypernuclei are very 
instructive, because their production is related to fundamental properties 
of the nucleation in nuclear matter \cite{SMM,Ogu11,Bot07,Buy18}. 
Therefore, the inclusion of larger nuclei in the analysis will allow to 
exclude alternative theoretical explanations. 

We specifically emphasize the needs for new experiments, and the 
measurement of new hypernuclei with different masses as 
extremely important for further progress in hypernuclear 
physics. The great variety of produced hypernuclei is an important 
advantage of relativistic heavy ion collisions in comparison with the 
traditional hypernuclear methods concentrated on reactions leading 
only to a few species. By comparing the yields of different hypernuclei 
produced in same reaction one can extract additional information 
about the properties of hypermatter, and also about exotic hypernuclei 
\cite{Buy18}. In this respect larger hypernuclear isotopes can provide 
further information (see also Ref.~\cite{Kit23}).

\section{Conclusions}

We have applied a novel theoretical approach to explain the yields of 
light nuclei and hypernuclei measured in the STAR experiment in 
the central relativistic nuclear collisions. 
Our approach combines 1) the adequate dynamical models for the baryon 
production in the first reaction stage, 2) the formation of intermediate 
local sources in equilibrium (excited large baryonic clusters) at subnuclear 
density, and 3) the description of the nucleation process inside these 
sources as their statistical decay. As was shown previously 
\cite{Bot21,Bot22} this approach can be successfully used 
to analyze the production of normal nuclei. We have demonstrated that 
the present experimental data concerning the production of hypernuclei 
can also be described within this approach by using 
similar parameters for the local sources. This may indicate an 
universal character of the nucleation process in rapidly expanding nuclear 
matter. 

We believe the production of hypernuclei in relativistic ion collisions 
opens the possibilities to 
investigate the properties of hypermatter at subnuclear densities, where 
the nuclei are formed. It is also important in astrophysics for models 
describing stellar matter in supernova explosions and in binary neutron 
star mergers. The suggested approach can correctly explain the 
main trends of the hypernuclear production, and provide consistent 
quantitative predictions too. We emphasized the puzzle 
of the relative yields of $^4_{\Lambda}$H and $^3_{\Lambda}$H nuclei in 
comparison with $^4$He and $^3$He yields, which can not 
be explained within the existing models, and which may carry information 
on the hyperon interaction during the formation of the hypernuclei. 
This can provide essential complementary information on the nuclear 
interaction, and on the nucleosynthesis process at low densities which was 
previously studied for normal nuclei only. Additional experiments on the 
comparative yields of several different hypernuclei in same reactions are 
crucially important. 

\begin{acknowledgments}
The authors acknowledges German Academic Exchange Service (DAAD) 
support from a PPP exchange grant and the Scientific and 
Technological Research Council of  T\"urkiye (TUBITAK) support 
under Project No. 121N420.  
T.R. acknowledges support through the Main-Campus-Doctus fellowship 
provided by the Stiftung Polytechnische Gesellschaft Frankfurt am Main 
and further thanks the Samson AG for their support. 
N.B. thanks J.W. Goethe University Frankfurt am Main for hospitality. 
Computational 
resources were provided by the Center for Scientific Computing (CSC) of 
the Goethe University and the "Green Cube" at GSI, Darmstadt. 
This publication is part of a project that has received funding from the 
European Union~s Horizon 2020 research and innovation programme 
under grant agreement STRONG -- 2020 -- No.824093. 
\end{acknowledgments}


\end{document}